\documentclass{iaus}
\usepackage{graphicx}

\title[Differential rotation on V889 Her] %% give here short title %%
{Differential rotation on the young solar analogue V889\,Herculis}

\author[Zs. K\H{o}v\'ari et al.]%% give here short author list %%
{Zsolt K\H{o}v\'ari$^1$,
Antonio Frasca$^2$,
Katia Biazzo$^3$,
Kriszti\'an Vida$^1$,
Ettore Marilli$^2$
\and \"Om\"ur \c{C}ak\i rl\i$^{4,}$$^5$}

\affiliation{$^1$Konkoly Observatory, \\
Budapest, XII. Konkoly Thege \'ut 15-17., H-1121, Hungary  \\ email: {\tt kovari@konkoly.hu}, {\tt vidakris@konkoly.hu} \\[\affilskip]
$^2$INAF, Osservatorio Astrofisico di Catania, \\
via S. Sofia, 78, 95123 Catania, Italy \\ email: {\tt antonio.frasca@oact.inaf.it},  {\tt ettore.marilli@oact.inaf.it}  \\[\affilskip]
$^3$INAF, Osservatorio Astrofisico di Arcetri, \\
L.go E. Fermi, 5, 50125 Firenze, Italy \\  email: {\tt kbiazzo@arcetri.astro.it}  \\[\affilskip]
$^4$Ege University, Science Faculty, Astronomy and Space Sciences Department,\\
35100 Bornova, Izmir, Turkey \\ email: {\tt omur.cakirli@ege.edu.tr}\\[\affilskip]
$^5$T\"UBITAK National Observatory, Akdeniz University Campus, 07058 Antalya, Turkey}

\pubyear{2010}
\volume{273}  %% insert here IAU Symposium No.
\pagerange{101--104}
\jname{Physics of Sun and Star Spots}
\editors{Debi Prasad Choudhary \& Klaus G. Strassmeier, eds.}
\begin{document}

\maketitle

\begin{abstract}
V889\,Herculis is one of the brightest single early-G type stars, a young Sun, that is rotating fast enough ($P_{\rm rot} =1.337$~days) for mapping its surface by Doppler Imaging. The 10 FOCES spectra collected between 13-16 Aug 2006 at Calar Alto Observatory allowed us to reconstruct one single Doppler image for two mapping lines. The Fe\,{\sc i}-6411 and Ca\,{\sc i}-6439 maps, in a good agreement, revealed an asymmetric polar cap and several weaker features at lower latitudes. Applying the sheared-image method with our Doppler reconstruction we perform an investigation to detect surface differential rotation (DR). The resulting DR parameter, $\delta\Omega/\Omega\approx0.009$ of solar type, is compared to previous studies which reported either much stronger shear or comparably weak DR, or just preferred rigid rotation. Theoretical aspects are also considered and discussed.

\keywords{stars: activity, stars: imaging, stars: individual (V889\,Her), stars: spots, stars: late-type}
\end{abstract}
\firstsection % if your document starts with a section,
\section{Introduction}
The primary source of all the manifestations of the solar activity is the magnetic dynamo.
However, the Sun is the only star, for which one of the conditions of a working dynamo,
the surface differential rotation can be measured directly, so far.
Today's advanced techniques can measure the solar surface and radial
rotation profiles with high accuracy, but at the dawn of extensive solar observations, tracing of
sunspots at different latitudes was the only tool to measure the surface shear on the Sun
(e.g. \cite[Maunder \& Maunder 1905]{2mau}). Similarly, on stars the detection of surface differential
rotation is still a challenging observational task. Direct tracing of
starspots is the usual way to observe surface shear, however, it requires
a reliable surface reconstruction technique such as Doppler imaging.

Detecting surface differential rotation (DR) on stars of different types can provide essential observational input
for the theoretical understanding of DR and dynamo.
Indeed, investigating solar analogues, such as our target, V889\,Her (HD\,171488), provide us with useful information
for further development of solar dynamo theory, which is still far from a
thorough understanding.

The fast-rotating solar-type V889\,Her is probably the brightest single early-G type
star that is rotating fast enough for mapping its photosphere by means of Doppler imaging. The star is a member of the Local Association,
a stream of young stars with ages ranging from 20 to 150 Myr (\cite[Montes et al. 2001]{Montes01}).
\cite[Strassmeier et al. (2003)]{Strassmeier03} made the first photometric and spectroscopic study of V889\,Her
and presented the first Doppler reconstruction showing a large polar spot and additional high-latitude features. Subsequent works by \cite[Marsden et al. (2006)]{Marsden06} and \cite[Jeffers \& Donati (2008)]{Jeffers08} based on Zeeman-Doppler imaging have also shown polar spottedness and found a strong solar-type differential rotation. However, \cite[J{\"a}rvinen et al. (2008)]{Jarvinen08} reported instead a much weaker surface shear, while \cite[Huber et al. (2009)]{Huber09} preferred rigid rotation.
In a comprehensive study of the chromospheric and photospheric activity, including Doppler imaging, \cite[Frasca et al. (2010)]{Frasca10} concluded that V889\,Her is a G2V type star. Hereafter, for further investigation of DR we adopt the reviewed astrophysical parameters given there (see Table~\ref{tab1}).

\begin{table}[t]
  \begin{center}
  \caption{Physical parameters of V889\,Her taken from \cite[Frasca et al. (2010)]{Frasca10}}
  \label{tab1}
  {\scriptsize
  \begin{tabular}{ll}\hline
  {\bf Parameter}                  & {\bf Value} \\  \hline
  Spectral type                   & G2V   \\
  $T_{\rm eff}$	(K)		  & 5750$\pm$130 \\
  $P_{\rm rot}$ (days)            & 1.33697$\pm$0.0002\\
  $v\sin i$ (km\,s$^{-1}$)        & 37.1$\pm$1.0   \\
  Inclination ($^{\circ}$)        & 60$\pm$10   \\
  $\log g$                        & $4.30\pm0.15$\\  
  Microturbulence (km\,s$^{-1}$)  & 1.6 \\
  Macroturbulence $\xi_{\rm R}=\xi_{\rm T}$ & 3.0 \\
  \hline
\end{tabular}
}
\end{center}
\end{table}

\begin{figure}[b]
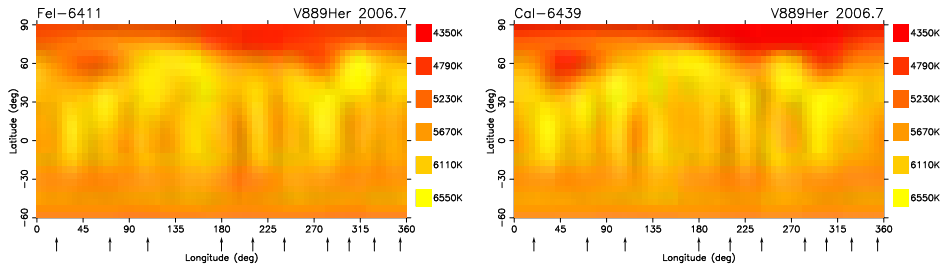

\begin{center}
%\hspace{-1cm}
\includegraphics[width=3.4cm, angle=-90]{kovari216_fig1a.ps}\hspace{0.3cm}\includegraphics[width=3.4cm, angle=-90]{kovari216_fig1b.ps}
\caption{Doppler maps for the  Fe\,{\sc i}-6411 lines ({\it right}) and for the Ca\,{\sc i}-6439 lines ({\it left}), taken from \cite[Frasca et al. (2010)]{Frasca10}. The arrows below indicate the phases of spectroscopic observations.} 
\label{fig:doppler_maps}
\end{center}
\end{figure}

\section{Sheared-image reconstruction}
In order to detect differential rotation, the Doppler reconstruction of \cite[Frasca et al. (2010)]{Frasca10} is repeated,
but instead of assuming rigid rotation, latitude-dependency is allowed in the rotation profile.
This technique known as 'sheared-image method' (\cite[Donati et al. 2000]{Donati00}) incorporates a predefined solar-type
differential rotation law in the Doppler imaging process. In practice, the image shear expressed by $\alpha=\delta\Omega/\Omega$ is
kept fixed, and individual Doppler reconstructions are done over a reasonable range of values in the $P_{\rm rot}-\alpha$ parameter plane.
Each of the resulting Doppler maps has a $\chi^2$-value which describes the goodness-of-fit to the observed spectral
line profiles. The best combination of $P_{\rm rot}$ and $\alpha$ is then marked by the least $\chi^2$ value.
The ability of this method was demonstrated also on artificial data by \cite[K\H{o}v\'ari \& Weber (2004)]{Kovariweb04}.

\begin{figure}[t]
\begin{center}
%\hspace{-1cm}
\includegraphics[width=5.8cm]{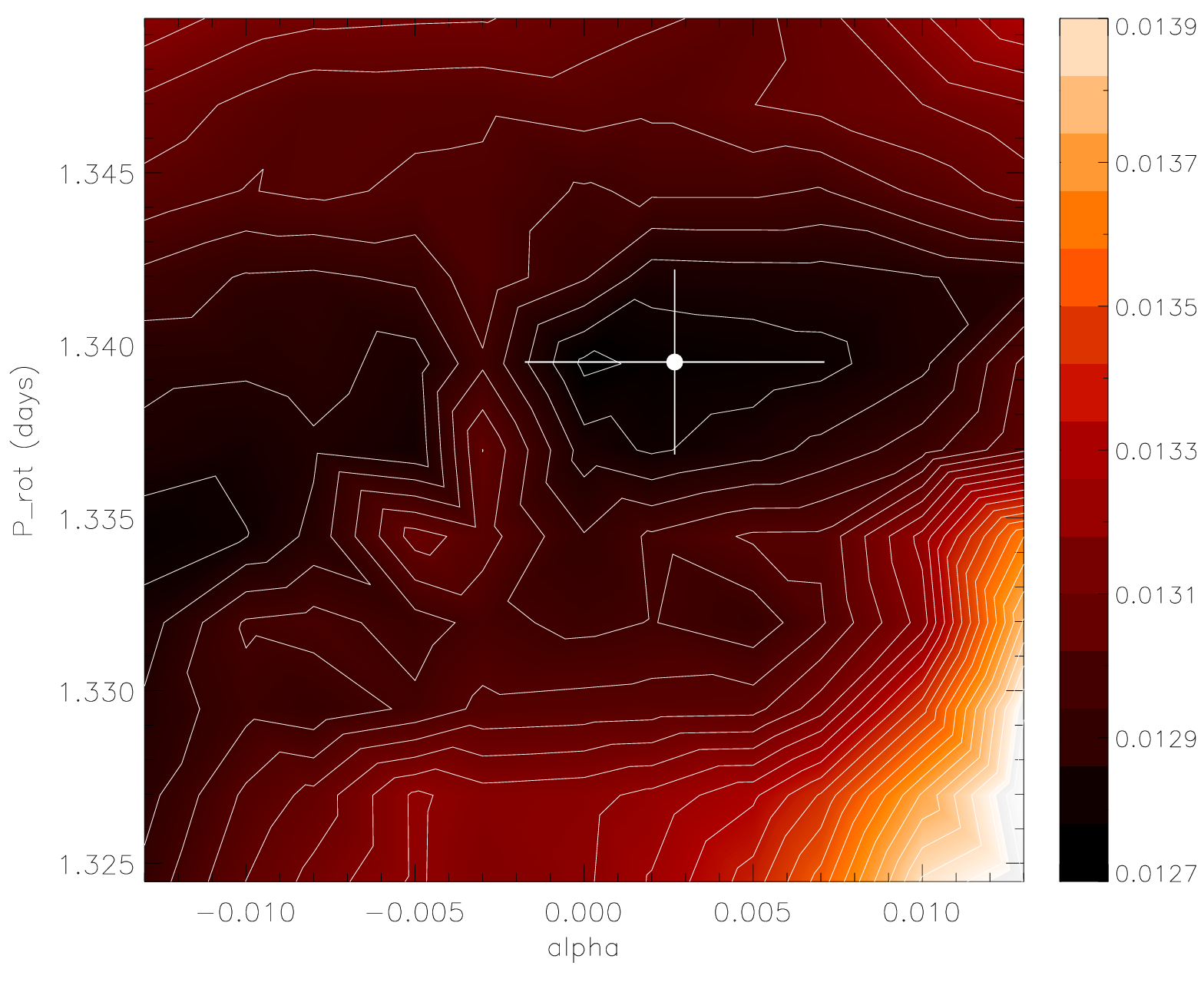}\hspace{0.6cm}\includegraphics[width=5.8cm]{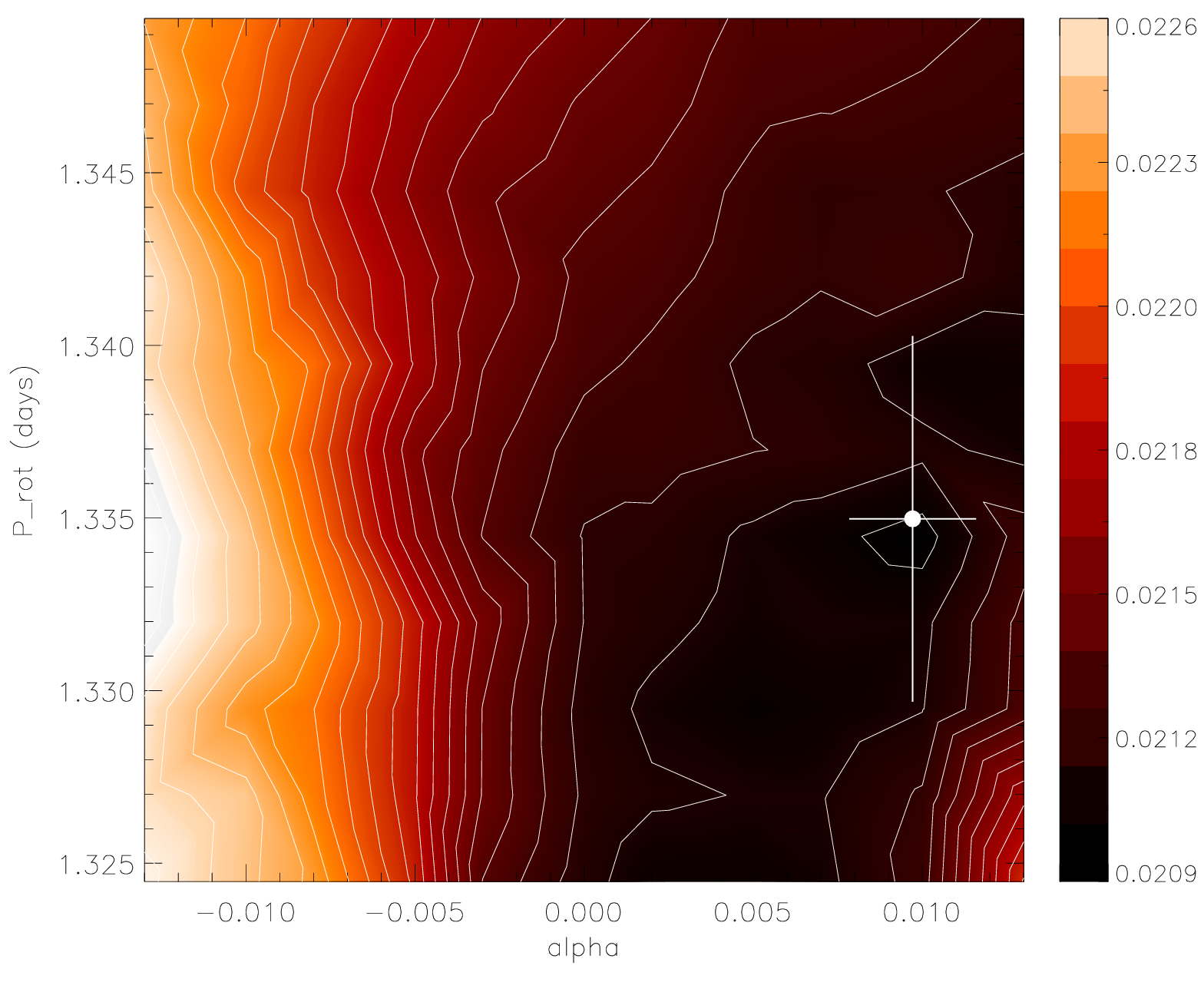}

\vspace{-0.2cm}
\includegraphics[width=5.8cm]{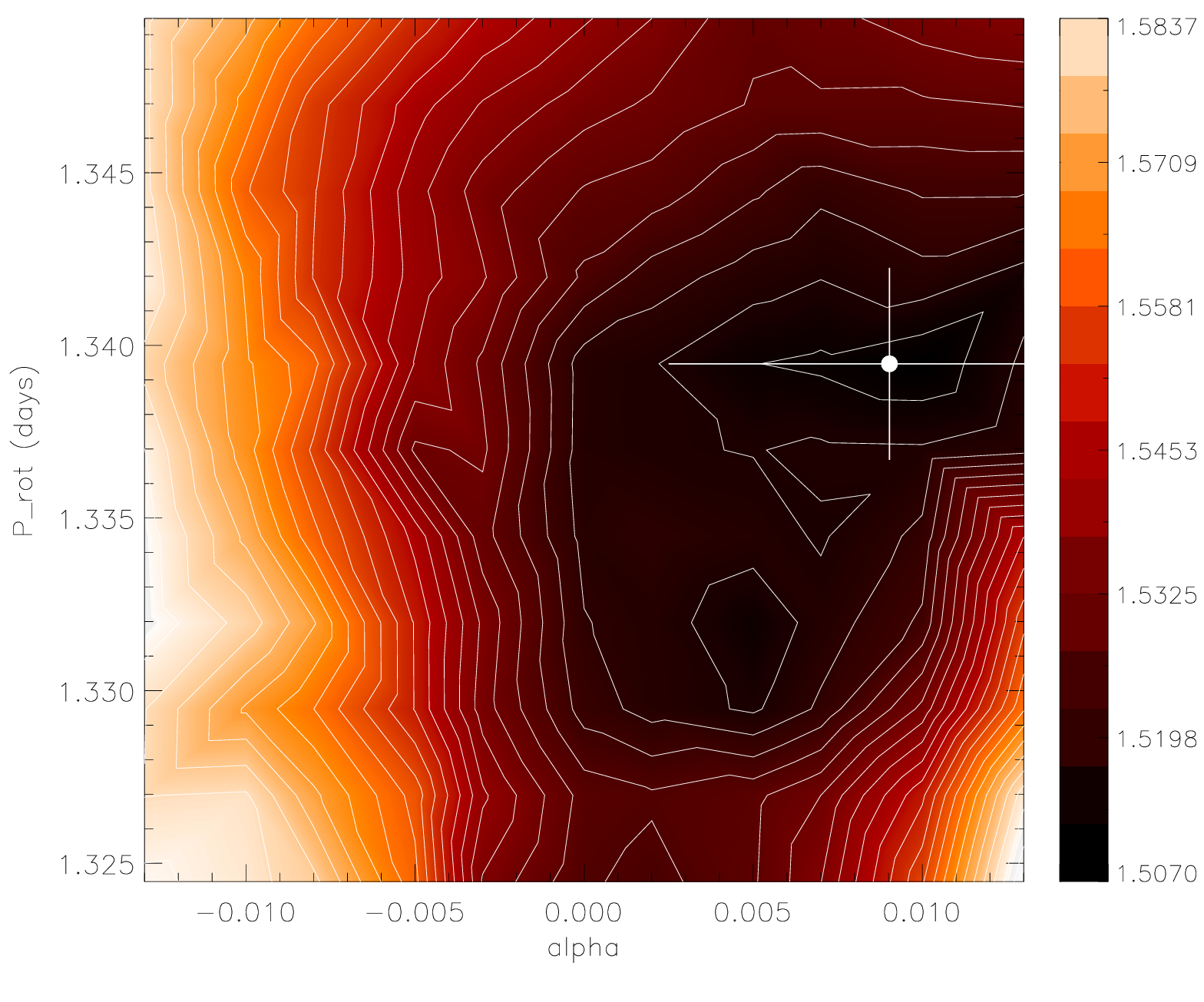}
\caption{Resulting $\chi^2$ maps of the sheared-image method over the $P_{\rm rot}-\alpha$ parameter plane. The best fit
is found at $\alpha=0.003\pm0.004$ and  $P_{\rm rot}=1.3395\pm0.0027$~days for the Fe\,{\sc i}-6411 reconstructions ({\it top left panel}) and at $\alpha=0.010\pm0.002$ and $P_{\rm rot}=1.3350\pm0.0052$~days for the Ca\,{\sc i}-6439 reconstructions ({\it top right panel}). Bottom panel shows the
averaged $\chi^2$ map, where minimum is found at $\alpha=0.009\pm0.006$ and at $P_{\rm rot}=1.3395\pm0.0028$~days.}
\label{chimaps}
\end{center}
\end{figure}

For the Doppler reconstructions we used the Doppler imaging code
{\sc TempMap} by \cite[Rice et al. (1989)]{Rice89}.
Doppler imaging was performed for the Fe\,{\sc i}-6411, 
and Ca\,{\sc i}-6439 mapping lines.
The individual reconstructions shown in Fig.~\ref{fig:doppler_maps} revealed similar spot distributions, i.e., mainly cool polar spots with temperature contrasts of up to $\approx$1500\,K with respect to the unspotted surface of $\approx$5800\,K.
Some low-latitude features are also recovered, however, with significantly weaker contrast ranging from $\approx$300\,K (Ca\,{\sc i}-6439)  
to a maximum of $\approx$500\,K (Fe\,{\sc i}-6411).

Fig.~\ref{chimaps} displays the $\chi^2$-maps of the trial-and-error
process over the  $P_{\rm rot}-\alpha$ parameter plane. The best fit
minima have similar locations for the two independent image reconstructions. We find
$\Omega (\theta) = 4.69 - 0.014 \sin^2 \theta$~rad/day
differential rotation law for the iron line and
$\Omega (\theta) = 4.71 - 0.047 \sin^2 \theta$~rad/day
for the calcium line, while the combined $\chi^2$ map suggests $\Omega_{\rm eq}= 4.69$~rad/day and $\delta\Omega= 0.042$~rad/day.

\section{Discussion}

Using the averaged $\chi^2$ landscape of the two independent line reconstructions in Fig.~\ref{chimaps} we find $\alpha=0.009$ or equivalently $\delta\Omega=\Omega_{\rm eq}-\Omega_{\rm pole}=0.042$~rad/day. Accordingly, the time the equator needs to lap the pole by one full
rotation is about 150~days, i.e., of the same order of the solar value. Observations and theoretical model calculations indicate that differential rotation is firmly influenced by stellar temperature (cf. e.g., \cite[Barnes et al. 2005]{Barnes2005}, \cite[Reiners 2006]{Reiners2006},
\cite[Kitchatinov \& R\"udiger 1995]{Kitrued95}).
A recent numerical model developed by \cite[Kitchatinov \& Olemskoy (2010)]{Kitol10} predicts $\delta\Omega\approx0.075$~rad/days for a $T_{\rm eff}=5800$\,K dwarf, which yields $\alpha\approx0.016$ at the angular velocity of V889\,Her. This shear is much weaker than $\delta\Omega\approx0.4-0.5$~rad/day obtained by \cite[Marsden et al. (2006)]{Marsden06} and by \cite[Jeffers \& Donati (2008)]{Jeffers08}, respectively, both from Zeeman Doppler imaging. 
Those values are among the largest ones measured only for
very few stars of mainly F-type (\cite[Reiners 2006]{Reiners2006}).
Such a large value does not fit the power law of \cite[Barnes et al. (2005)]{Barnes2005} which suggests $\delta\Omega\approx0.12$~rad/day for a G2-type star. Moreover, \cite[J{\"a}rvinen et al. (2008)]{Jarvinen08} argues for a substantially weaker differential rotation.
Indeed, \cite[Huber et al. (2009)]{Huber09} preferred solid rotation, though, due to the rather low data quality, their
results does not exclude a weak surface shear such as that was suggested by \cite[J{\"a}rvinen et al. (2008)]{Jarvinen08} and by us.

All in all, the differential rotation derived from our sheared-image process for the G2-star V889\,Her is below but close to the one from
the empirical relationship in \cite[Barnes et al. (2005)]{Barnes2005} and seems to support the theoretical model predictions (\cite[K\"uker \& R\"udiger 2005]{Kurud05}, \cite[Kitchatinov \& Olemskoy 2010]{Kitol10}) as well.
On the other hand, such a weak differential rotation is inconsistent with the high surface shear derived from Zeeman Doppler imaging (\cite[Marsden et al. 2006]{Marsden06}, \cite[Jeffers \& Donati 2008]{Jeffers08}). This difference might partially be explained by temporal variations in differential rotation, however, in the case of V889\,Her \cite[Jeffers et al. (2010)]{Jeffers10} found no evidence for such a temporal change.
If so, other effects such as rapid spot evolution should also be considered.
In addition, the very different methods and data applied to detect the surface shear could have different, sometimes depreciated or unrecognized error sources, which can also explain such a disagreement (see also T.A. Carroll's paper in this volume). For instance, when various spectral features used for imaging correspond to different surface layers, the detected shear would diverge.

\begin{acknowledgments}
ZsK and KV are supported by the Hungarian Science Research Program (OTKA) grants K-68626 and K-81421.
This work has been supported by the Italian {\em Ministero dell'Istruzione, Universit\`a e  Ricerca} (MIUR) and by the 
{\em Regione Sicilia}.
\end{acknowledgments}

\end{document}